# Intense Internal and External Fluorescence as Solar Cells Approach the Shockley-Queisser Efficiency Limit

Owen D. Miller and Eli Yablonovitch

Material Sciences Division, Lawrence Berkeley National Laboratory, Berkeley, CA 94720

Electrical Engineering & Comp Sciences Dept., University of California, Berkeley, CA 94720

Sarah R. Kurtz

National Renewable Energy Laboratory, Golden, CO 80401

## ABSTRACT:

Absorbed sunlight in a solar cell produces electrons and holes. But, at the open circuit condition, the carriers have no place to go. They build up in density and, ideally, they emit external fluorescence that exactly balances the incoming sunlight. Any additional non-radiative recombination impairs the carrier density buildup, limiting the open-circuit voltage. At open-circuit, efficient external fluorescence is an indicator of low internal optical losses. Thus efficient external fluorescence is, counter-intuitively, a necessity for approaching the Shockley-Queisser efficiency limit. A great Solar Cell also needs to be a great Light Emitting Diode.

Owing to the narrow escape cone for light, efficient external emission requires repeated attempts, and demands an internal luminescence efficiency >>90%.

## Introduction

The Shockley-Queisser (SQ) efficiency limit [1] for a single junction solar cell is ~33.5% under the standard AM1.5G flat-plate solar spectrum [2]. In fact, detailed calculations in this paper show that GaAs is capable of achieving this efficiency. Nonetheless, the record GaAs solar cell had achieved only 26.4% efficiency [3] in 2010. Previously, the record had been 26.1% [4] and prior to that stuck [5] at 25.1%, during 1990-2007. Why then the 7% discrepancy between the theoretical limit 33.5% versus the previously achieved efficiency of 26.4%?

It is usual to blame material quality. But in the case of GaAs double heterostructures, the material is almost ideal [6] with an internal fluorescence yield of >99%. This deepens the puzzle as to why the full theoretical SQ efficiency is not achieved?

## The Physics Required to Approach the Shockley-Queisser Limit

Solar cell materials are often evaluated on the basis of two properties: how strongly they absorb light, and whether the created charge carriers reach the electrical contacts, successfully. Indeed, the short-circuit current in the solar cell is determined entirely by those two factors. However, the power output of the cell is determined by the product of the current and voltage, and it is therefore imperative to understand what material properties (and solar cell geometries) produce high voltages. We show here that maximizing the external emission of photons from the front surface of the solar cell proves to be the key to reaching the highest possible voltages. In the search for optimal solar cell candidates, then, materials that are good radiators, in addition to being good absorbers, are most likely to reach high efficiencies.

As solar efficiency begins to approach the SQ limit, the internal physics of a solar cell transforms. Shockley and Queisser showed that high solar efficiency is accompanied by a high concentration of carriers, and by strong fluorescent emission of photons. In a good solar cell, the photons that are emitted internally are likely to be trapped, re-absorbed, and re-emitted, leading to “photon recycling” at open-circuit.

The SQ limit assumes perfect external fluorescence yield at open-circuit. On the other hand, inefficient external fluorescence at open-circuit is an indicator of non-radiative recombination and optical losses. Owing to the narrow escape cone, efficient external emission requires repeated escape attempts, and demands an internal luminescence efficiency >>90%. We find that the failure to efficiently extract the recycled internal photons is an indicator of an accumulation of non-radiative losses, which are largely responsible for the failure to achieve the SQ limit in the best solar cells.

In high efficiency solar cells it is important to engineer the photon dynamics. The SQ limit requires 100% external fluorescence to balance the incoming sunlight at open circuit. Indeed, the external fluorescence is a thermodynamic measure [7] of the available open-circuit voltage. Owing to the narrow escape cone for internal photons, they find it hard to escape through the semiconductor surface. Thus external fluorescence efficiency is always significantly lower than internal fluorescence efficiency. Then the SQ limit is not achieved.

The extraction and escape of internal photons is now recognized as one of the most pressing problems in light emitting diodes (LED’s). In this article, we assert that luminescence extraction is *equally* important to solar cells. **The Shockley-Queisser limit cannot be achieved unless light extraction physics is designed into high performance solar cells, which requires that non-radiative losses be minimized, just as in LED’s.**

In some way this is counter-intuitive, since an extracted photon cannot contribute to performance. Paradoxically, 100% external extraction at open-circuit is exactly what is needed to achieve the SQ limit. The paradox is resolved by recognizing that high extraction efficiency at open-circuit is an indicator, or a gauge, of small optical losses. Previous record solar cells have typically taken no account of light extraction. **Nonetheless, achieving the 33.5% SQ limit will require light extraction to become part of all future designs.** The present shortfall below the SQ limit can be overcome.

Figure 1: The drastic effect of internal fluorescence efficiency, $\eta_{int}$, on theoretical solar cell efficiency. The shortfall is particularly noticeable for smaller bandgaps. A reduction from $\eta_{int}$=100% to $\eta_{int}$=90% already causes a large drop in performance, while a reduction from $\eta_{int}$=90% to $\eta_{int}$=80% causes little additional damage. Owing to the need for photon recycling, and the multiple attempts required to escape the solar cell, $\eta_{int}$ must be >>90%.

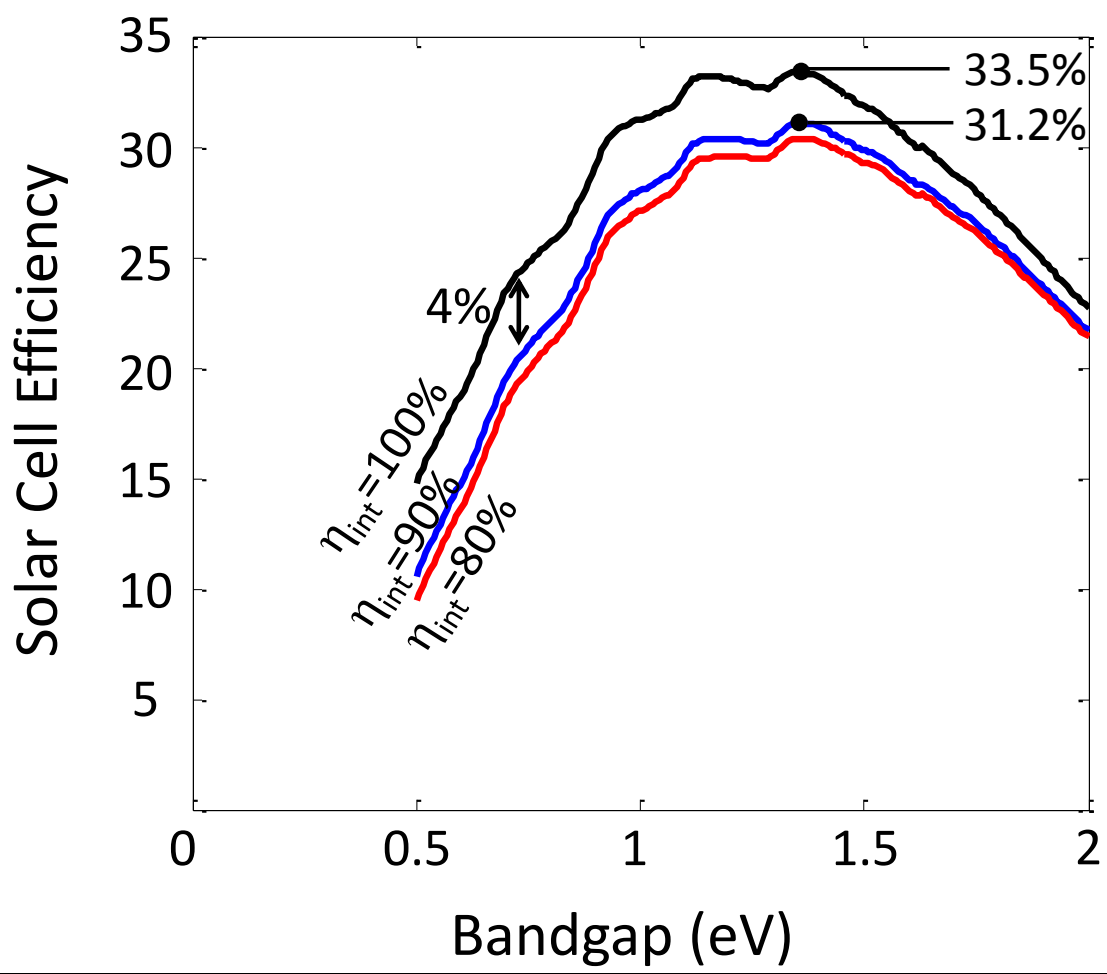

Although Silicon makes an excellent solar cell, its internal fluorescence [8,9] yield is <20%, which prevents Silicon from approaching the SQ limit. The physical issues presented here pertain to any material which has the possibility of approaching the SQ limit, which requires near unity external fluorescence as III-V materials can provide, and that perhaps other material systems can provide as well.

Figure 2: The drastic effect of rear mirror reflectivity on cell efficiency and on open-circuit voltage, $V_{oc}$., but not on short circuit current $J_{sc}$, for a 3μm thick GaAs solar cell. Mirror reflectivity >>90% makes a big difference, owing to the small escape cone for external emission, and the multiple attempts needed for escape.

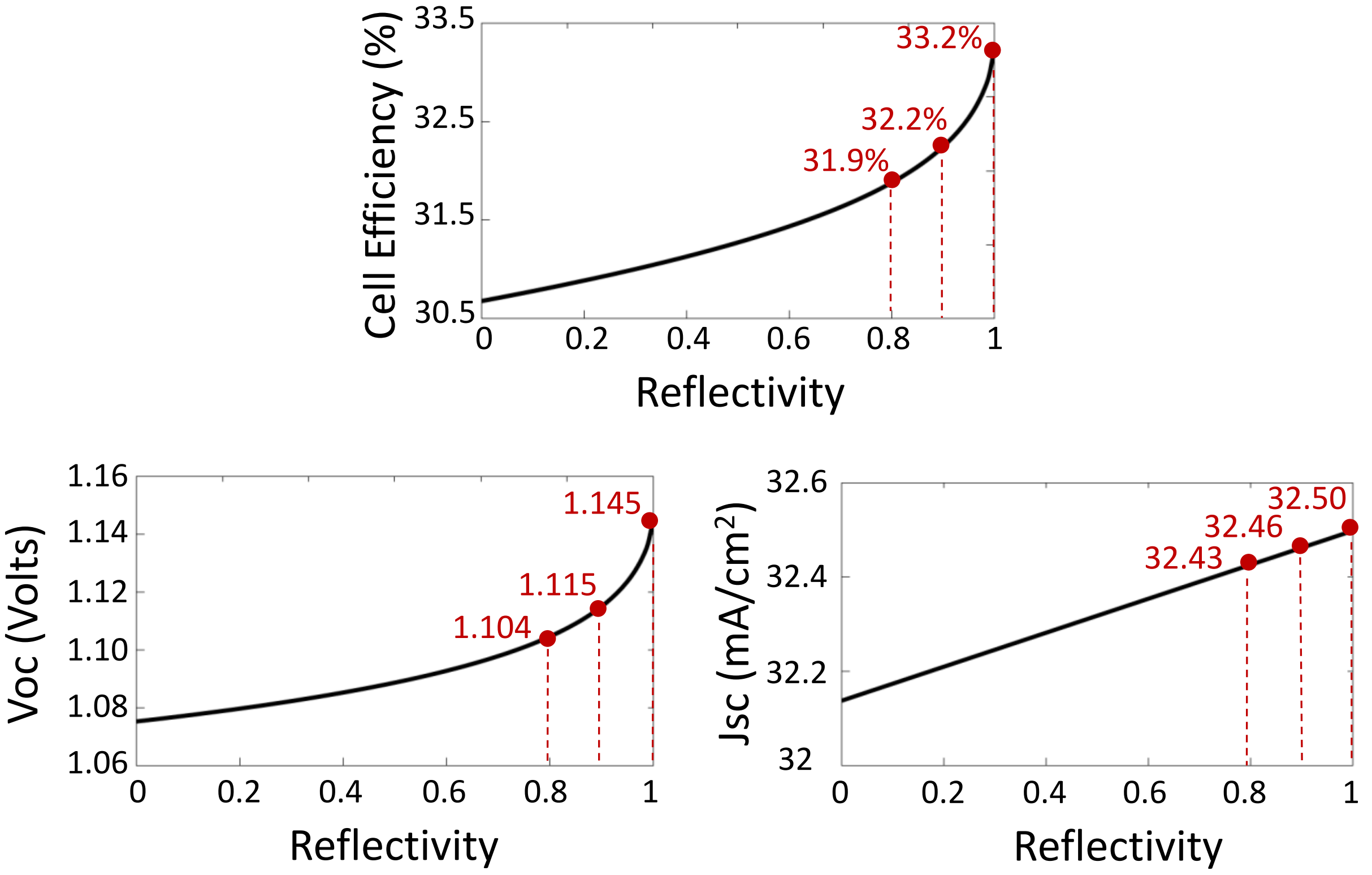

Since light is trapped by total internal reflection, it is likely to be re-absorbed, leading to a further re-emission event. With each absorption/re-emission event, the solid angle of the escape cone [10] allows only ($1/4n^2$) ~ 2% of the internal light to escape. This puts a very heavy burden on the parasitic losses in the cell. With only 2% escaping per emission event, even a 90% internal fluorescence yield on each cycle would appear inadequate. Likewise the rear mirror should have >>90% reflectivity. This is illustrated in Figures 1 and 2.

A good solar cell should be designed as a good light emitting diode, with good light extraction. In a way, this is not surprising. Most ideal machines work by reciprocity, equally well in reverse. This has important ramifications. For ideal materials the burden of high open-circuit voltage, and thereby high efficiency, lies with optical design: The solar cell must be designed for optimal light extraction under open-circuit conditions.

The assumption of perfect internal fluorescence yield is a seductive one. The Shockley-Queisser limit gets a significant boost from the perfect photon recycling that occurs in an ideal system. Unfortunately, for most materials, their relatively low internal fluorescence yields mean that the upper bounds on their efficiencies are much lower than the Shockley-Queisser limit. For the few material systems that are nearly ideal, such as GaAs, there is still a tremendous burden on the optical design of the solar cell. A very good rear mirror, for example, is of the utmost importance. In addition, it becomes clear that realistic material radiative efficiencies must be included in a credible assessment of any materials' prospects as a solar cell technology.

There is a well-known detailed balance equation relating the spontaneous emission rate of a semiconductor to its absorption coefficient [11]. Nevertheless, it is not true that all good absorbers are good emitters. If the non-radiative recombination rate is higher than the radiative rate then the probability of emission will be very low. Amorphous silicon, for example, has a very large absorption coefficient of about $10^5$/cm, yet the probability of emission at open circuit is ~$10^{-4}$. The probability of internal emission in high-quality GaAs has been experimentally tested to be 99.7% [6]. GaAs is a unique material in that it both absorbs and radiates well, enabling the high voltages required to reach >30% efficiency.

The idea that increasing light emission improves open-circuit voltage seems paradoxical, as it is tempting to equate light emission with loss. Basic thermodynamics dictates that materials which absorb sunlight must also emit in proportion to their absorptivity. Thus electron-hole recombination producing external fluorescent emission is a necessity in solar cells. At open circuit, external photon emission is part of a necessary and unavoidable equilibration process, which does not represent loss at all.

At open circuit an ideal solar cell would in fact radiate out of the solar cell a photon for every photon that was absorbed. Any *additional* non-radiative recombination, or photon loss, would indeed waste photons and electrons. Thus the external fluorescence efficiency is a gauge or an indicator of whether the additional loss mechanisms are present. In the case of no additional loss mechanisms, we can look forward to 100% external fluorescence, and maximum open circuit voltage, $V_{oc}$. At the power-optimized, solar cell operating bias point [12], the voltage is slightly reduced, and 98% of the open-circuit photons are drawn out of the cell as real current. Good external extraction comes at no penalty in current at the operating bias point.

On thermodynamic grounds, Ross [7] had already proposed the open circuit voltage would be penalized by poor external fluorescence efficiency $\eta_{ext}$ as:

$$V_{oc} = V_{oc-ideal} - kT\left|\ln \eta_{ext}\right| \quad (1)$$

This can be derived as follows: Under ideal open-circuit, quasi-equilibrium conditions, the solar pump rate equals the external radiative rate: $R_{ext} = P_{pump}$. If the radiative rate is diminished by a poor external fluorescence efficiency $\eta_{ext}$, the remaining photons must have been wasted in non-radiative recombination or parasitic optical absorption. The effective solar pump is then reduced to $P_{pump}\times\eta_{ext}$. The quasi-equilibrium condition is then $R_{ext}=P_{pump}\times\eta_{ext}$ at open circuit. Since the radiative rate $R_{ext}$ depends on the carrier density $np$ product, which is proportional to $\exp\{qV_{oc}/kT\}$, then the poor extraction $\eta_{ext}$ penalizes $V_{oc}$ just as indicated in Eqn. (1).

Another way of looking at this is to notice the shorter carrier lifetime in the presence of the additional non-radiative recombination. We start with a definition $\eta_{ext}\equiv R_{ext}/(R_{ext} + R_{nr})$, where $R_{nr}$ is the internal photon and carrier non-radiative loss rate per unit area. Simple algebraic manipulation shows that the total loss rate ($R_{ext} + R_{nr}$) = $R_{ext}/\eta_{ext}$. Thus a poor $\eta_{ext}$<1 increases the total loss rate in inverse proportion, and the shorter lifetime limits the build-up of carrier density at open circuit. Then carrier density is connected to exp{$qV_{oc}/kT$} as before.

It is important to emphasize that light emission from *only* the front surface of the solar cell should be maximized. Solar photons are only incident on the front, so the only unavoidable balancing emission is that of fluorescent photons exiting through the front. Having a good mirror on the rear surface greatly improves the photon recycling mechanism and therefore the voltage.

## Theoretical Efficiency Limits of GaAs Solar Cells

The Shockley-Queisser limit includes a major role for external fluorescence from solar cells. Accordingly, internal fluorescence followed by light extraction, plays a direct role in determining theoretical efficiency. To understand these physical effects a specific material system must be analyzed, replacing the hypothetical step function absorber stipulated by SQ.

GaAs is a good material example, where external fluorescence extraction plays an important role in determining the fundamental efficiency prospects. The quasi-equilibrium approach developed by SQ [1] is the most rigorous method for calculating such efficiency limits. Properly adapted, it can account for the precise incoming solar radiation spectrum, the real material absorption spectrum, the internal fluorescence efficiency, as well as the external extraction efficiency and light trapping [10]. Calculations including such effects for silicon solar cells were completed more than 25 years ago [13,14]. Surprisingly, a calculation with the same sophistication has not yet been completed for GaAs solar cells.

Previous GaAs calculations have approximated the solar spectrum to be a blackbody at 6000°K, and/or the absorption coefficient to be a step function [1,15,16]. The efficiency limits calculated with these assumptions are all less than or equal to 31%.

In this paper we calculate that the theoretical maximum efficiency of a GaAs solar cell, using [2] the one-sun AM1.5G solar spectrum, and the proper absorption curve of GaAs, is in fact 33.5%. Allowing for practical limitations, it should be possible to manufacture flat-plate single-junction GaAs solar cells with efficiencies above 30% in the near future. As we have already shown, realizing such efficiencies will require optical design such that the solar cell achieves optimal light extraction at open circuit.

Consider a solar cell in the dark surrounded by a thermal bath at room temperature $T$. The surrounding environment radiates at $T$ according to the tail ($E>>kT$) of the blackbody formula:

$$b(E) = \frac{2n_r^2 E^2}{h^3 c^2} \exp(-E/kT) \tag{2}$$

where $b$ is given in photons per unit area, per unit time, per unit energy, per steradian. $E$ is the photon energy, $n_r$ is the ambient refractive index, $c$ is the light speed, and $h$ is Planck's constant. As Lambertian distributed photons enter the solar cell's surface at polar angle $\theta$, with energy $E$, the probability they will be absorbed is written as the dimensionless absorbance $a(E,\theta)$. The flux per unit solid angle of absorbed photons is therefore $a(E,\theta)b(E)$. In thermal equilibrium there must be an emitted photon for every absorbed photon; the flux of emitted photons per unit solid angle is then also $a(E,\theta)b(E)$.

When the cell is irradiated by the sun, the system will no longer be in thermal equilibrium. There will be a chemical potential separation, $\mu$, between electron and hole quasi-Fermi levels. The emission spectrum, which depends on electrons and holes coming together, will be multiplied by the normalized *np* product, $(np/n_i^2)$, where $n$, $p$, and $n_i$ are the excited electron and hole concentration, and the intrinsic carrier concentration, respectively. The Law of Mass Action is $np=n_i^2\exp\{\mu/kT\}$ for the excited semiconductor in quasi-equilibrium, where $\mu$ is the internal chemical potential created by the sunlight. Then, the total photon emission rate is:

$$R_{ext} = \exp\{\mu / kT\} \times \int\int a(E,\theta) b(E)\,\mathrm{d}E \cos\theta\,\mathrm{d}\Omega \tag{3}$$

for external solid angle Ω and polar angle θ. Eqn. (3) is normalized to the flat plate area of the solar cell, meaning that the emission rate $R_{ext}$ is the emissive flux from only the front surface of the solar cell. There will generally be a much larger photon flux inside the cell, but most of the photons undergo total internal reflection upon reaching the semiconductor-air interface. If the rear surface is open to the air, i.e. there is no mirror, then the rear surface emission rate will equal the front surface emission rate. As already discussed, restricting the luminescent emission to the front surface of the solar cell improves voltage, whereas a faulty rear mirror increases the avoidable losses, significantly reducing the voltage.

To find the open-circuit voltage we now equate the carrier recombination and generation rates. Carriers are generated by the incident solar radiation *S(E)* according to the formula:

$$P_{pump} = \int\int a(E,\theta) S(E)\,\mathrm{d}E \cos\theta\,\mathrm{d}\Omega \tag{4}$$

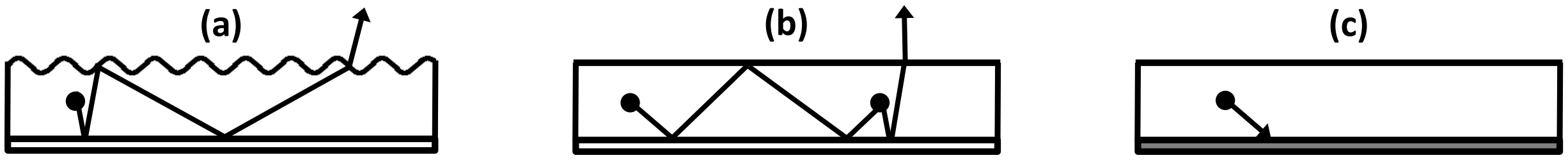

Figure 3: Solar performance from three geometries: (a) randomly textured front surface with a perfectly reflecting mirror on the rear, (b) planar front surface with a perfectly reflecting mirror on the rear surface, and (c) planar front surface with an absorbing mirror on the rear surface.

Equating the generation and recombination rates, and recognizing that the open-circuit voltage equals the quasi-Fermi level separation ($qV_{OC} = \mu$), the resulting open-circuit voltage is:

$$V_{oc} = \frac{kT}{q} \ln\left\{ \frac{\int\int a(E,\theta)\, S(E) dE \cos\theta\, d\Omega}{\int\int a(E,\theta) b(E)\, dE \cos\theta d\Omega} \right\} - \frac{kT}{q} |\ln\{\eta_{ext}\}| \quad (5)$$

which is an expanded version of Eqn. (1). Since $\eta_{\text{ext}}$ is ≤1, the second term in Eqn. (5) represents a loss of voltage due to poor light extraction.

To explore the physics of light extraction, we consider GaAs solar cells with three possible geometries, as shown in Fig. 3. The first geometry, 3(a), is the most ideal, with a randomly textured front surface and a perfectly reflecting mirror on the rear surface. The surface texturing enhances absorption and improves light extraction, while the mirror ensures that the photons exit from the front surface and not the rear. The second geometry, 3(b), uses a planar front surface while retaining the perfectly reflecting mirror. Finally, the third geometry Fig. 3(c) has a planar front surface and an absorbing rear mirror, which captures most of the internally emitted photons before they can exit the front surface. We will show that this configuration achieves almost the same short-circuit current as the others, but suffers greatly in voltage and, consequently, efficiency. Thus the optical design affects the voltage more than it does the current. Please note that the 3(c) geometry is equivalent to the common situation in which the active layer is epitaxially grown on top of an electrically passive substrate, which absorbs without re-emission.

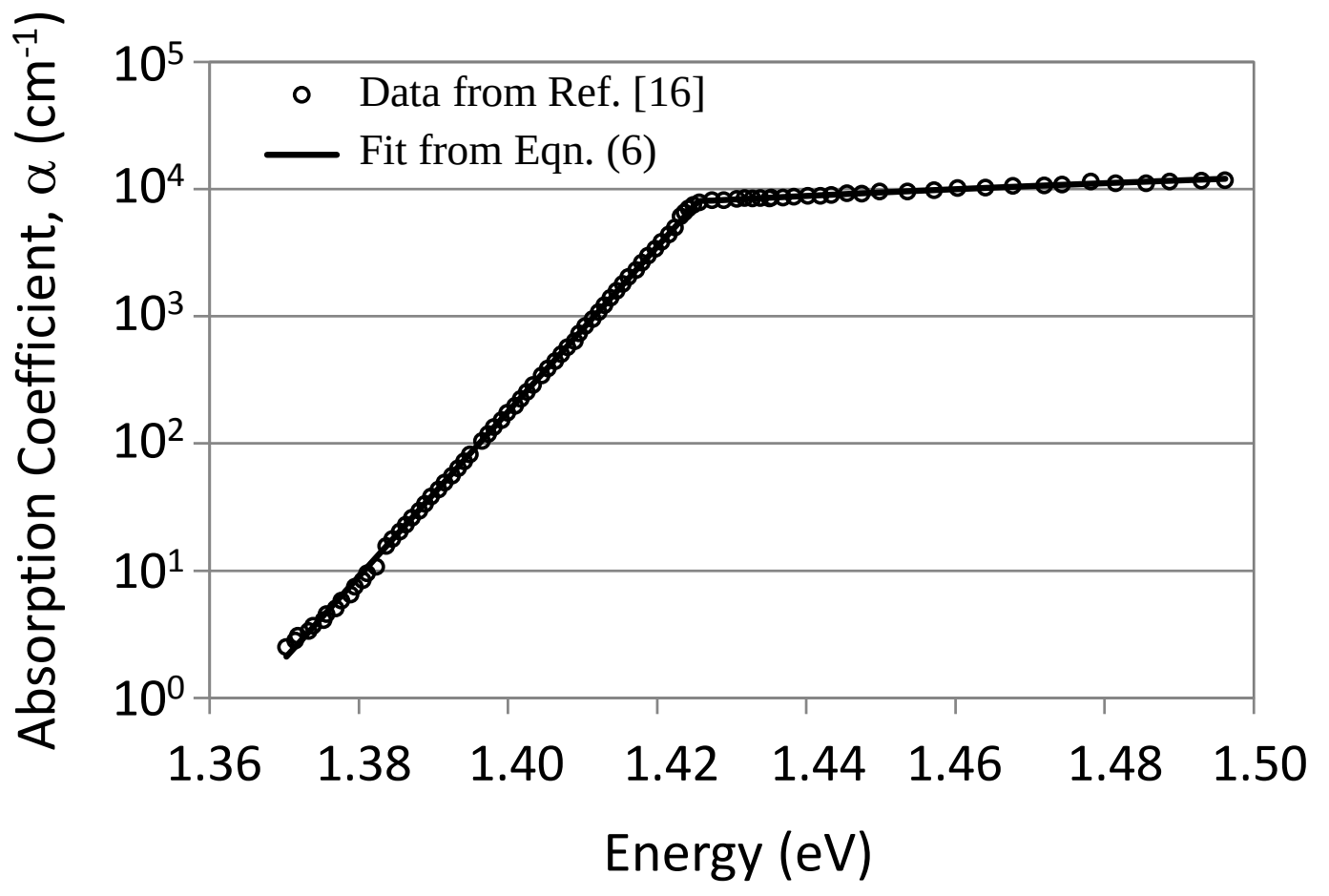

Figure 4: GaAs absorption coefficient as a function of energy. GaAs has a direct bandgap at $E_g$=1.42 eV, with an “Urbach tail” that falls off exponentially for lower energies.

GaAs has a 1.4 eV bandgap that is ideally suited for solar cells. It is a direct-bandgap material, with an absorption coefficient of 8000 $cm^{-1}$ near its band-edge. By contrast, the absorption coefficient of Si is ~$10^4$ times weaker at its indirect band-edge. Fig. 4 is a semi-log plot of the absorption coefficient as a function of energy; the circles are experimental data from [17] while the solid line is a fit to the data using the piecewise continuous function:

$$\alpha = \begin{cases} \alpha_o \exp\left(\dfrac{E - E_g}{E_o}\right) & E \le E_g \\ \alpha_o \left(1 + \dfrac{E - E_g}{E'}\right) & E > E_g \end{cases} \tag{6}$$

where $\alpha_o$ = 8000/cm, the Urbach energy is $E_o$ = 6.7meV, and $E'$ = 140meV. The exponential dependence of the absorption coefficient below the bandgap is characteristic of the "Urbach tail" [18].

Efficient external emission can be separated into two steps: First, the semiconductor should have a substantially higher probability of recombining radiatively, rather than non-radiatively. We define the internal fluorescence yield $\eta_{int}$, similarly to the external fluorescence yield, as the probability of radiative recombination versus non-radiative recombination, $\eta_{int} \equiv R_{int}/(R_{int} + R_{nr})$, where $R_{int}$ and $R_{nr}$ are the radiative and non-radiative recombination rates per unit volume, respectively. The internal fluorescence yield is a measure of intrinsic material quality. The second factor for efficient emission is proper optical design, to ensure that the internally radiated photons eventually make their way out to external surface of the cell. Maximizing both factors is crucial for high open-circuit and operating point voltages.

We now derive the external fluorescence yield for the three different geometries. At open-circuit, $P_{pump}$ and the recombination rates, ($R_{ext}$ + $R_{nr}$), are equal, and this allowed the derivation of Eqn. (5) for a general open-circuit voltage. In operation, however, current will be drawn from the solar cell and the two rates will not be equal. The current will be the difference between pump and recombination terms:

$$J = q\left(P_{pump} - R_{ext} - R_{nr}\right) = \int\int a(E,\theta)S(E)\,\mathrm{d}E\cos\theta\,\mathrm{d}\Omega - \frac{1}{\eta_{ext}}q\pi e^{qV/kT}\int_0^\infty a(E)b(E)\mathrm{d}E \tag{7}$$

where the external luminescence from the cell is a Lambertian that integrates to $\pi$ steradians, and the absorption $a(E,\theta)$ has been assumed independent of polar angle $\theta$, which is clearly valid for the randomly textured surface. It is independent of incident angle for a planar front surface because the large refractive index of GaAs refracts the incident light very close to perpendicular inside the solar cell.

Case (a): The Randomly Textured Surface:

Randomly texturing the front surface of the solar cell, Fig. 3(a), represents an ideal method for coupling incident light to the full internal optical phase space of the cell. The absorption of a textured cell has been derived in Ref. [19]:

$$a(E) = \frac{4n^2\alpha L}{4n^2\alpha L + 1} \tag{8}$$

Although only strictly valid in the weakly absorbing limit, the absorptivity is close enough to one for large $\alpha L$ that Eqn. (8) can be used for all energies.

To derive the external fluorescence yield, all of the recombination mechanisms must be identified. We have assumed a perfectly reflecting rear mirror, so the net radiative recombination is the emission from the front surface, given by Eqn. (3). The only fundamental non-radiative loss mechanism in GaAs is Auger recombination. The Auger recombination rate per unit area is $CLn_i^3\exp\{3qV/2kT\}$, where $L$ is the thickness of the cell, $C$=7x10$^{-30}$ cm$^6$s$^{-1}$ is [20] the Auger coefficient, and intrinsic doping is assumed, to minimize the Auger recombination. The external fluorescence yield can then be written:

$$\eta_{ext}(V) = \frac{\pi e^{qV/kT}\int_0^\infty a(E)b(E)\mathrm{d}E}{\pi e^{qV/kT}\int_0^\infty a(E)b(E)\mathrm{d}E + CLn_i^3 e^{3qV/2kT}} \tag{9}$$

Case (b): Planar Front Surface with Perfectly Reflecting Mirror:

A second interesting configuration to consider is that of Fig. 3(b), which has a planar front surface and a perfectly reflecting rear mirror. Comparison with the first configuration allows for explicit determination of the improvement introduced by random surface texturing. Not surprisingly, surface texturing only helps for very thin cells.

The absorptivity of the planar cell is well-known:

$$a(E) = 1 - e^{-2\alpha L} \tag{10}$$

where the optical path length is doubled because of the rear mirror. Using this absorptivity formula, Eqn. (3) still represents the external emission. As a consequence, the external fluorescence yield follows the same formula, Eq'n. (9), albeit with different absorptivities, $a(E)$, for the planar front surface versus the textured solar cell.

Case (c): Planar Front Surface with Absorbing Mirror:

We have emphasized the importance of light extraction at open circuit to achieve a high voltage. To demonstrate the effects of poor optical design on efficiency, we also consider the geometry of Fig. 3(c). No extra recombination mechanism has been introduced, but the rear mirror now absorbs light rather than reflecting it internally. (Or equivalently, it transmits light into a non-radiating, optically lossy, substrate.)

To calculate the external fluorescence yield of this geometry, one could explicitly calculate the probability of internally emitted light escaping. However, a simpler approach is to realize that the geometry with an absorbing rear-mirror is equivalent to a setup with an absorbing non-fluorescent substrate supporting the active material, as depicted in Fig. 5. Viewed either way, the absorptivity is: $a(E) = 1 - e^{-\alpha L}$, where the light now has the opportunity for only one pass through the semiconductor, to become absorbed.

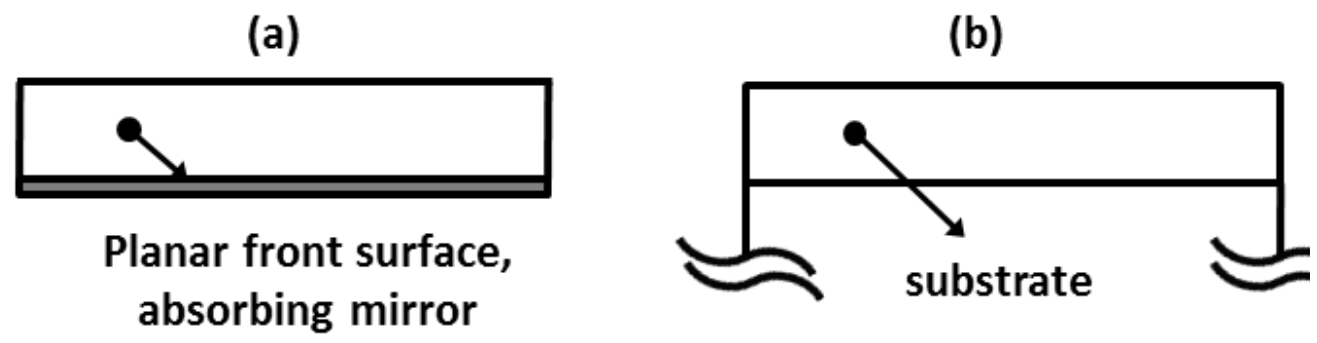

Figure 5: Optically equivalent configurations, assuming the substrate does not reflect, nor re-radiate, absorbed photons. In both cases, almost all of the internally emitted light will be lost out of the rear surface of the solar cell.

To calculate the external fluorescence yield, one can use the rate balancing method described earlier. The recombination terms for emission out of the front surface and Auger processes are still present. Now there is also a term for emission out of the rear surface. By the same reasoning as for front surface emission, the emission out of the rear surface balances the thermal radiation coming from below: $a'(E,\theta')\times b'(E)\times\exp\{qV/kT\}$ which includes a further boost by the quasi-equilibrium factor $\exp\{qV/kT\}$. At the rear surface, the density of states of the internal blackbody radiation $b'(E)\equiv n_r^2 b(E)$ is increased by $n_r^2$, where $n_r$ is the refractive index of the semiconductor. The rear absorption $a'(E,\theta)$ is also modified as shown in the following equation for the total number of incident photons absorbed per unit area:

$$2\pi n_r^2 \int_0^\infty b(E) \int_0^{\pi/2} \left(1 - e^{-\frac{f(\theta')\alpha L}{\cos\theta'}}\right) \cos\theta' \sin\theta' \mathrm{d}\theta' \quad \text{where } f(\theta') = \begin{cases} 1 & \theta' < \theta_c \\ 2 & \theta' > \theta_c \end{cases} \tag{11}$$

where the $2\pi$ prefactor arises from the azimuthal integral, and $f(\theta')$ accounts for the different path lengths traveled by internal photons at angles greater or less than the critical angle $\theta_c$, defined by the escape cone at the top surface. The internal path length by oblique rays is increased by the factor $1/\cos\theta'$. A similar expression for the rear absorption is found in [16]. The external fluorescence yield is now the ratio of the emission out of the front surface to the sum of the emission out of either surface plus Auger recombination:

$$\eta_{ext}(V) = \frac{\pi e^{qV/kT}\int_0^\infty a(E)b(E)\mathrm{d}E}{\pi e^{qV/kT}\int_0^\infty b(E)\left[a(E) + 2n_r^2\int_0^{\pi/2}\left(1 - e^{-\frac{f(\theta')\alpha L}{\cos\theta'}}\right)\sin\theta'\cos\theta'\mathrm{d}\theta'\right]\mathrm{d}E + CLn_i^3 e^{3qV/2kT}} \tag{12}$$

which is an explicit function of the quasi-Fermi level separation $V$.

Given the absorptivity and external fluorescence yield of each geometry, calculation of the solar cell's I-V curve and power conversion efficiency is straightforward using Eqn. (7). The power output of the cell, $P$, is simply the current multiplied by the voltage. The operating point (i.e. the point of maximum efficiency) is the point at which $dP/dV = 0$. Substituting the absorption coefficient data and solar spectrum values into Eqn. (7), it is simple to numerically evaluate the bias point where the derivative of the output power equals zero.

Figure 6 is a plot of the solar cell efficiencies as a function of thickness for the three solar cell configurations considered. Also included is a horizontal line representing the best GaAs solar cell fabricated up to 2010, which had an efficiency of 26.4% [3]. The maximum theoretical efficiency is 33.5%, more than 7% larger in absolute efficiency. An efficiency of 33.5% is theoretically achievable for both planar and textured front surfaces, provided there is a mirrored rear surface.

Figure 6: GaAs solar cell efficiency as a function of thickness. Random surface texturing does not increase the limiting efficiency of 33.5%, although it enables high efficiencies even for cell thickness less than one micron. Having an absorbing mirror on the rear surface incurs a voltage penalty and reduces the theoretical limiting efficiency to 31.1%. There is still a sizeable gap between the 26.4% cell and the theoretical limit. The cell thickness was not specified in Ref. [3].

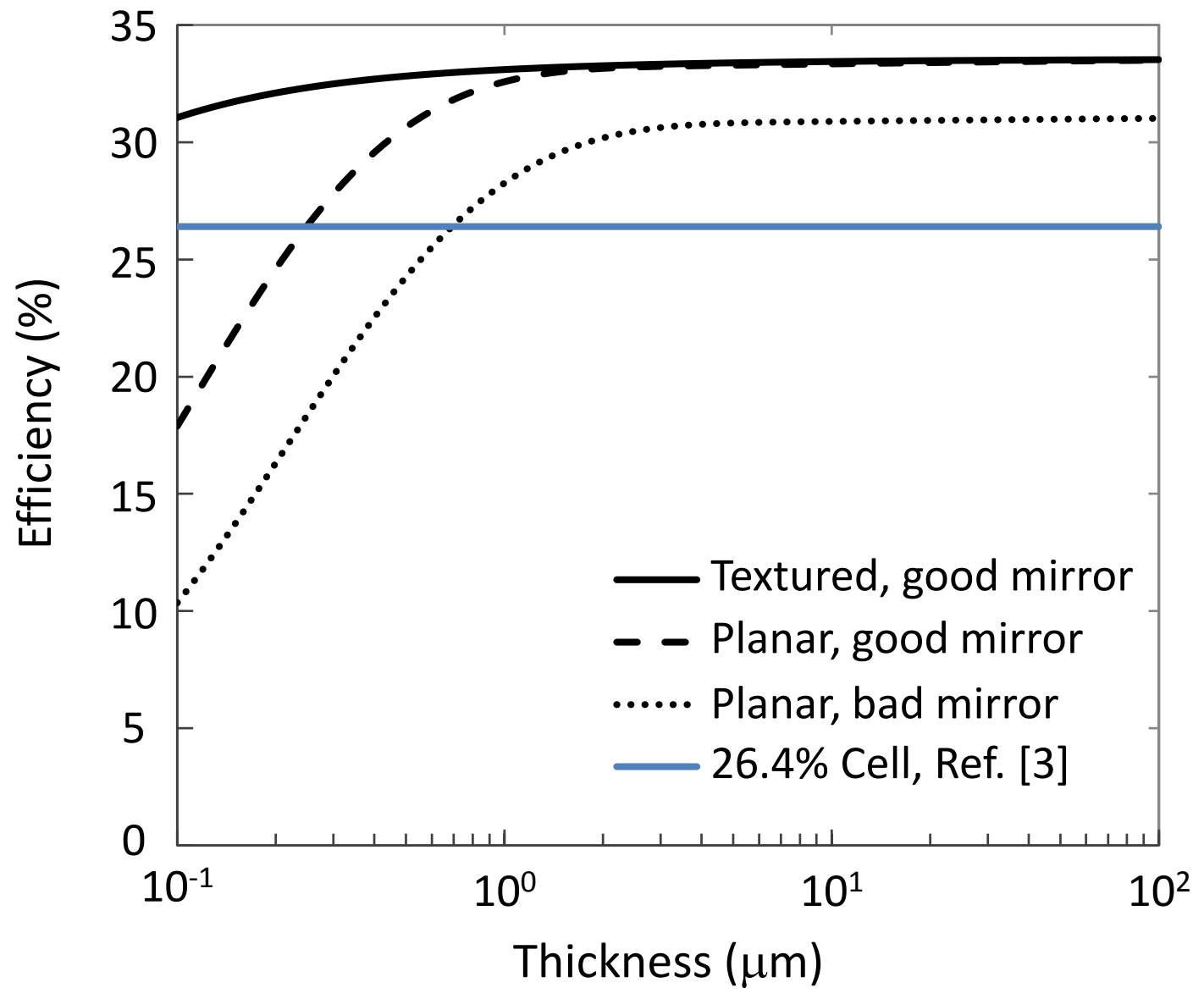

Although surface texturing does not increase the maximum efficiency, it does help maintain an efficiency greater than 30% even for solar cells that are only a few hundred nanometers thick. The cell with a planar surface and bad mirror on its rear surface reaches an efficiency limit of only 31.1%, exhibiting the penalty associated with poor light extraction. To understand more clearly the differences that arise in each of the three configurations, the short-circuit currents and open-circuit voltages of each are plotted in Figure 7.

Table 1 and Figure 7 display the differences in performance between a planar solar cell with a perfect mirror and one with an absorbing mirror. Although the short-circuit currents are almost identical for both mirror qualities, at thicknesses greater than 2-3μm, the voltage differences are drastic. Instead of reflecting photons back into the cell where they can be re-absorbed, the absorbing mirror constantly removes photons from the system. The photon recycling process attendant to a high external fluorescence yield is almost halted when the mirror is highly absorbing.

Figure 7: (a) Open-circuit voltage ($V_{OC}$), and (b) short-circuit current ($J_{SC}$), as a function of thickness for each of the solar cell configurations considered. The planar cell with an absorbing mirror reaches almost the same short-circuit current as the other two configurations, but it suffers a severe voltage penalty due to poor light extraction and therefore lost photon recycling. There is considerable opportunity to increase $V_{oc}$ over the previous record 26.4% cell. (The textured cell/good mirror has a lower voltage than a planar cell/good mirror owing to the effective bandgap shift observed in Fig. 10. This slight voltage drop is not due to poor $\eta_{ext}$.)

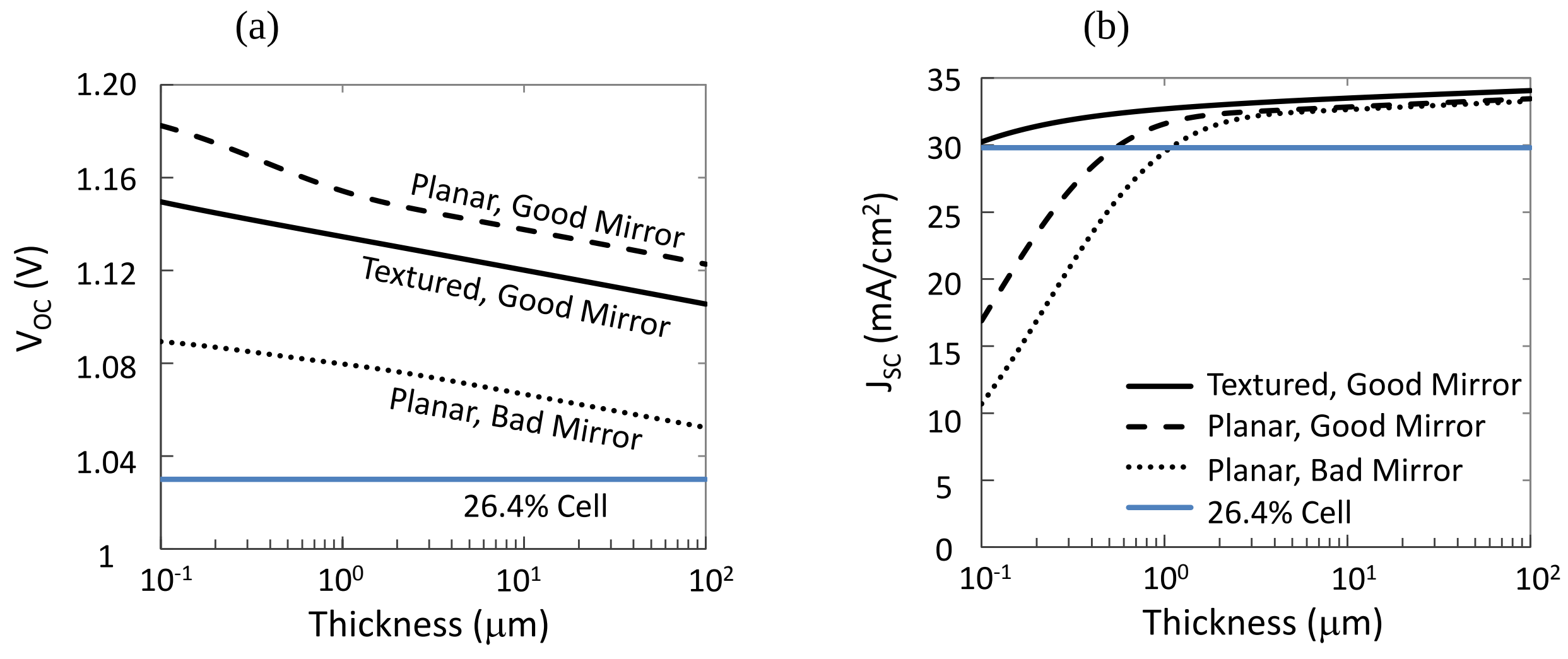

Fig. 8 presents such intuition visually, displaying the internal and external currents of a 10μm thick GaAs solar cell at its maximum power point, for 0% and 100% reflectivity. In both cases, cells absorb very well, and the short-circuit currents are almost identical. The extracted currents, too, are almost identical. In each case 0.8mA of current is “lost”; (i.e. the difference between short-circuit and operating currents), but in the case of the good mirror the current is lost to front surface emission. There is a strong buildup of photon density when the only loss is emission through the front surface, allowing much higher internal luminescence and carrier density. A higher operating voltage results.

Table 1: Table of $V_{oc}$, $J_{sc}$, and efficiency values for three possible geometries and relevant cell thicknesses. A good rear mirror is crucial to a high open-circuit voltage, and consequently to efficiencies above 30%.

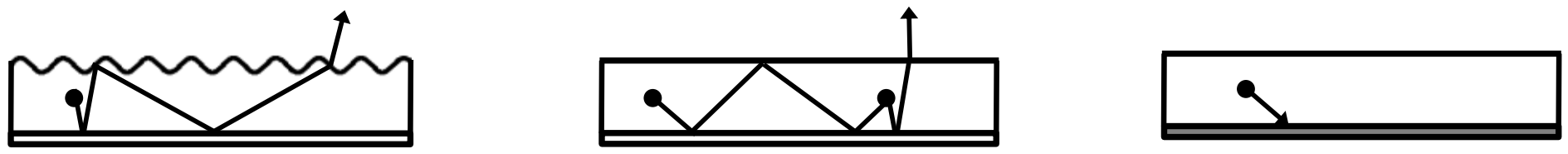

| | **Textured good mirror** | | | **Untextured good mirror** | | | **Untextured bad mirror** | | |
|---|---|---|---|---|---|---|---|---|---|
| Thickness | 500nm | 1 µm | 10µm | 500nm | 1µm | 10µm | 500nm | 1µm | 10µm |
| Voc (volts) | 1.14 | 1.13 | 1.12 | 1.16 | 1.15 | 1.14 | 1.08 | 1.08 | 1.07 |
| Jsc (mA/cm$^2$) | 32.3 | 32.7 | 33.5 | 29.5 | 31.6 | 32.8 | 25.2 | 29.5 | 32.6 |
| Fill Factor | 0.89 | 0.89 | 0.89 | 0.89 | 0.89 | 0.89 | 0.89 | 0.89 | 0.89 |
| efficiency % | 32.8 | 33.1 | 33.4 | 30.6 | 32.6 | 33.3 | 24.3 | 28.3 | 30.9 |

From Fig. 6, it is clear that surface texturing is not helpful in GaAs, except to increase current in the very thinnest solar cells. In most solar cells, such as silicon cells, surface texturing provides a mechanism for exploiting the full internal optical phase space. The incident sunlight is refracted into a very small solid angle within the cell, and without randomizing reflections, photons would never couple to other internal modes.

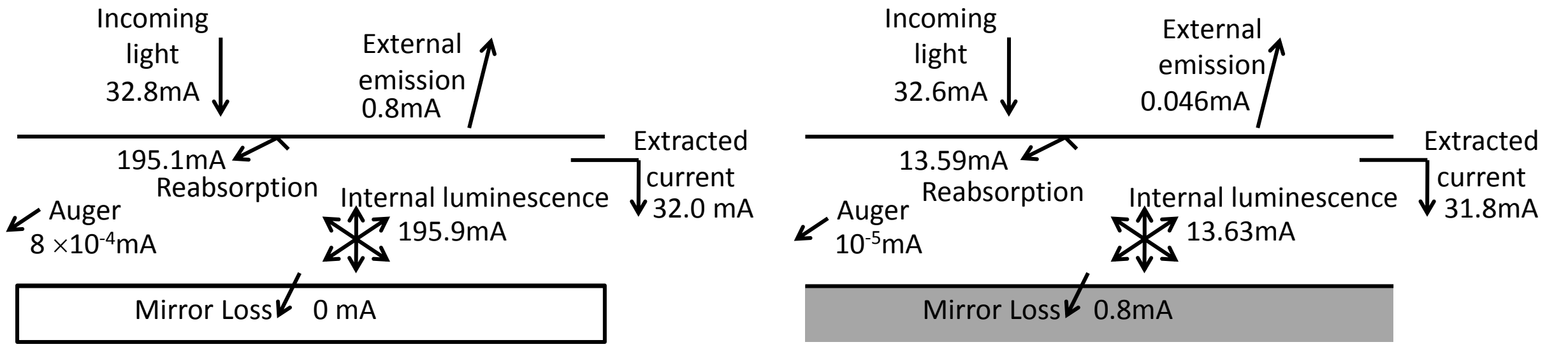

Figure 8: Diagram of currents at the operating point for a planar 10µm thick, 1cm$^2$ solar cell (a) with a perfectly reflecting mirror and (b) with an absorbing mirror. The two solar cells produce almost the same amount of current, but the cell with a perfectly reflecting mirror achieves a higher voltage. The internal luminescence and re-absorption demonstrate the impact of photon recycling on carrier densities. In GaAs, Auger recombination is negligible.

Figure 9: Qualitative illustration of the different photon dynamics in plane-parallel solar cells with (a) low fluorescence yield $\eta_{ext}$ and (b) high $\eta_{ext}$, respectively. In (a), the lack of photon recycling reduces carrier density and external emission. Conversely in (b), the internal photons achieve full angular randomization even without surface texturing, and the high external emission is indicative of high carrier density build-up through photon recycling.

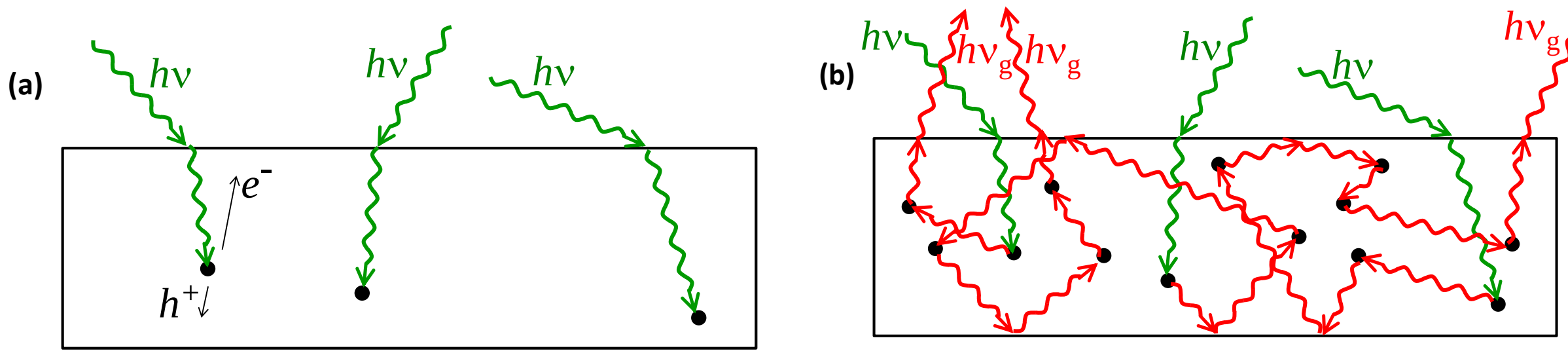

GaAs is such an efficient radiator that it can provide the angular randomization by photon recycling. After absorbing a photon, the photon will likely be re-emitted, and the re-emission is equally probable into all internal modes. Whereas most materials require surface roughness to efficiently extract light, the radiative efficiency of GaAs ensures light extraction based on photon recycling. Such photon dynamics are illustrated in Figure 9.

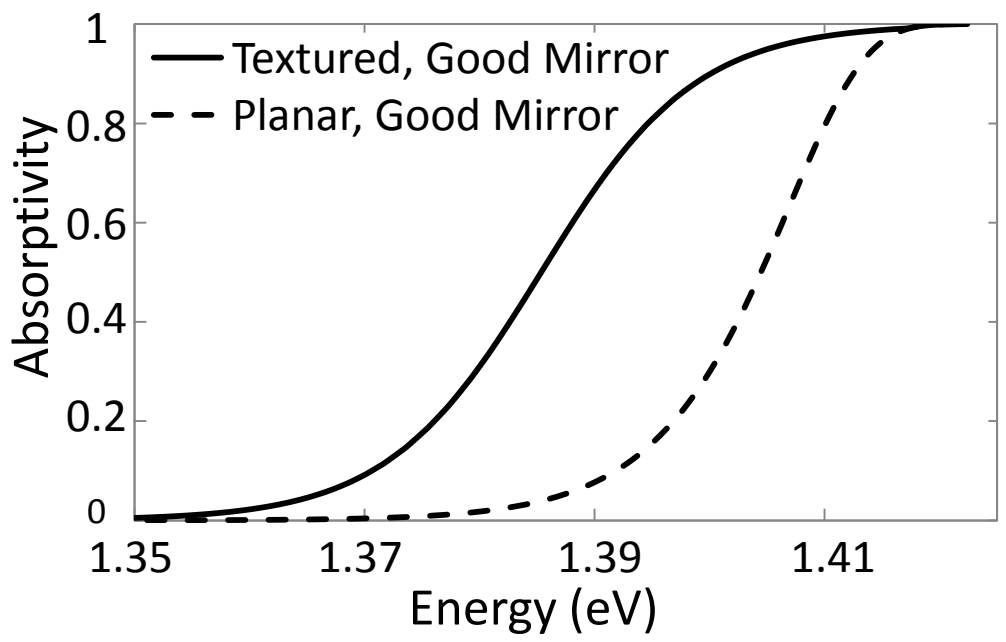

Figure 10: Absorptivity of a 10μm thick cell as a function of photon energy near the bandgap. The solid line represents a textured front surface, while the dotted line represents a plane parallel slab. A randomly textured solar cell absorbs well even below 1.39eV, effectively shifting the bandgap to a lower energy and lowering $V_{oc}$.

Marti [16] already emphasized the benefits of photon recycling toward efficiency, but we believe that external fluorescence yield is the more comprehensive parameter for boosting solar cell efficiency and voltage.

It seems surprising that the planar solar cell would have a higher voltage than the textured cell. This is due to a second-order effect seen in Fig. 10. Textured cells experience high absorption even below the bandgap, due to the longer optical path length provided. Texturing effectively reduces the bandgap slightly, as shown in Fig. 10, accounting for the lower open-circuit voltage but larger short-circuit current values seen in Fig. 7(b).

**Discussion:**

GaAs is an example of one of the very few material systems that can reach internal fluorescence yields close to 1; a value of 99.7% has been experimentally confirmed [6]. However, it is the *external* fluorescence yield that determines voltage, and that yield depends on both the quality of material and also the optical design. Absorbing contacts, or a faulty rear mirror, for example, will remove photons from the system that could otherwise be recycled. Additionally an optically textured design [10] can provide the possibility for extraction of luminescent photons, before they could be lost.

Light trapping is normally thought of as a way to absorb more light and increase the current in a thinner cell. But the concentration of carriers in a thinner cell also provides a voltage increase, $\Delta V \sim kT\ln\{4n^2\}$, an effect that was implicitly used when light trapping was first incorporated into the fundamental calculation [13] of Silicon efficiency . Thus texturing improves the voltage in most solar cells. Nonetheless, one of the main results of this article is that the voltage boost can come with *OR* without surface texturing in GaAs. The reason is that efficient internal photon recycling provides the angular randomization necessary to concentrate the light, even in a plane-parallel GaAs cell. Thus, short-circuit current in GaAs can benefit from texturing, but GaAs voltage accrues the same benefit with, *OR* even without, texturing.

The distinction between voltage boost by texturing, and voltage boost by photon recycling was already made by Lush and Lundstrom [21] who predicted the higher voltages and the record efficiencies that have recently been [22] observed in thin film III-V solar cells. However, the over-arching viewpoint in this article is that voltage is determined by external fluorescence efficiency. That viewpoint accounts in a single comprehensive manner for the benefits of nano-texturing, photon re-cycling, parasitic optical reflectivity, and imperfect fluorescence, while being thermodynamically self-consistent.

In the case of perfect photon recycling, there is surprisingly little thickness dependence of $V_{oc}$. This is to be contrasted with the textured case, where the voltage boost might require light concentration and carrier concentration within a thin cell. Under perfect photon re-cycling, photons are lost only at the surface, and the photon density and carrier density are maintained at the maximum value through the full depth. The solar cell can be permitted to become thick, with no penalty. In practice a thick cell would carry a burden, and an optimum thickness would emerge.

**Conclusions:**

We have shown how to include photon recycling and imperfect radiation properties into the quasi-equilibrium formulation of Shockley and Queisser. High voltages $V_{oc}$ are achieved by maximizing the external fluorescence yield of a system. Using the standard solar spectrum and the measured absorption curve of GaAs, we have shown that the theoretical efficiency limit of GaAs is 33.5%, which is more than 4% higher than that of silicon [23], and achieves its efficiency in a cell that is 100 times thinner.

Internally trapped radiation is necessary, but not sufficient, for the high external luminescence that allows a cell to reach voltages near the theoretical limits. The optical design must ensure that the only loss mechanism is photons exiting at the front surface. A slightly faulty mirror, or equivalently absorbing contacts or some other optical loss mechanism, sharply reduces the efficiency limit that can be achieved. To realize solar cells with efficiency greater than 30%, the optical configuration will need to be very carefully designed.

The prior [3] efficiency record, 26.4%, was set by GaAs cells that had $V_{oc}$=1.03Volts. Alta Devices has recently [22] made a big improvement in GaAs efficiency and open-circuit voltage, 28.2% and $V_{oc}$=1.11Volts respectively, showing in part the benefit of light extraction.

The Shockley-Queisser formulation is still the foundation of solar cell technology. However, the physics of light extraction and external fluorescence yield are clearly relevant for high performance cells and will prove important in the eventual determination of which solar cell technology wins out in the end. In the push for high-efficiency solar cells, a combination of high-quality GaAs and optimal optical design should enable single-junction flat plate solar cells with greater than 30% efficiency.

## Acknowledgements

This work was supported by the Department of Energy ‘Light-Material Interactions in Energy Conversion’ Energy Frontier Research Center under grant DE-SC0001293. O.D.M. acknowledges the National Science Foundation for fellowship support.

REFERENCES

1. W. Shockley and H.J. Queisser, “Detailed Balance Limit of p-n Junction Solar Cells,” Journal of Applied Physics 32, 510 (1961).

2. National Renewable Energy Laboratory, ASTM G-173-03, Reference Solar Spectral Irradiance: Air Mass 1.5, website: http://rredc.nrel.gov/solar/spectra/am1.5/

3. M.A. Green, K. Emery, Y. Hishikawa, and W. Warta, “Solar cell efficiency tables (version 36),” Prog. Photovolt: Res. Appl. 18, 346 (2010).

4. G.J. Bauhuis, P. Mulder, E.J. Haverkamp, J.C.C.M. Huijben, and J.J. Schermer, "26.1% thin-film GaAs solar cell using epitaxial lift-off," Solar Energy Materials & Solar Cells 93, 1488 (2009)

5. M.A. Green, K. Emery, Y. Hishikawa, and W. Warta, “Solar cell efficiency tables (version 31),” Prog. Photovolt: Res. Appl. 16, 61 (2008).

6. I. Schnitzer, E. Yablonovitch, C. Caneau, and T.J. Gmitter, “Ultrahigh spontaneous emission quantum efficiency, 99.7% internally and 72% externally, from AlGaAs/GaAs/AlGaAs double heterostructures,” Applied Physics Letters 62, 2 (1993).

7. R.T. Ross, “Some Thermodynamics of Photochemical Systems,” Journal of Chemical Physics 46, 4590 (1967).

8. E. Yablonovitch and T. Gmitter, “Auger Recombination in Silicon At Low Carrier Densities,” Applied Physics Letters 49, 587 (1986).

9. T. Trupke, J. Zhao, A. Wang, R. Corkish, and M.A. Green, “Very Efficient Light Emission from Bulk Crystalline Silicon,” Applied Physics Letters 82, 2996 (2003).

10. E. Yablonovitch and G.D. Cody, “Intensity Enhancement in Textured Optical Sheets for Solar Cells,” IEEE Transactions on Electron Devices 29, 300 (1982).

11. W. van Roosbroeck and W. Shockley, “Photon-Radiative Recombination of Electrons and Holes in Germanium,” Physical Review 94, 1558 (1954).

12. P. Wurfel, *Physics of solar cells,* Wiley-VCH, p. 124, (1995).

13. T. Tiedje, E. Yablonovitch, G.D. Cody, and B.G. Brooks, “Limiting Efficiency Of Silicon Solar Cells”, IEEE Transactions on Electron Devices 31, 711 (1984).

14. P. Campbell and M.A. Green, “The Limiting Efficiency of Silicon Solar Cells under Concentrated Sunlight,” IEEE Transactions on Electron Devices 33, 234 (1986).

15. J.L. Balenzategui and A. Marti, “The Losses of Efficiency in a Solar Cell Step By Step,” 14$^{th}$ E.U. Photovoltaic Solar Energy Conference, 2374 (1997).

16. A. Marti, J.L. Balenzategui, and R.F. Reyna, “Photon recycling and Shockley’s diode equation,” Journal of Applied Physics 82, 4067 (1997).

17. M.D. Sturge, “Optical Absorption of Gallium Arsenide between 0.6 and 2.75 eV,” Physical Review 127, 768 (1962).

18. F. Urbach, “The Long-Wavelength Edge of Photographic Sensitivity and of the Electronic Absorption of Solids,” Physical Review 92, 1324 (1953).

19. E. Yablonovitch, “Statistical Ray Optics,” J. Opt. Soc. Am. 72, 899 (1982).

20. U. Strauss, W.W. Ruhle, and K. Kohler, “Auger recombination in intrinsic GaAs,” Applied Physics Letters 62, 55 (1993).

21. G. Lush and M. Lundstrom, “Thin Film Approaches for High Efficiency III-V Cells”, Solar Cells **30**, 337 (1991).

22. M.A. Green, K. Emery, Y. Hishikawa, W. Warta, & E.D. Dunlop, “Solar cell efficiency tables (version 38),” Prog. Photovolt: Res. Appl. 19, 565 (2011).

23. M.J. Kerr, A. Cuevas, and P. Campbell, “Limiting Efficiency of Crystalline Silicon Solar Cells Due to Coulomb-Enhanced Auger Recombination,” Prog.Photovolt: Res. Appl. 11, 97 (2003).